\documentclass[12pt]{article}
\begin{document}
\newcommand{\be}{\begin{equation}}
\newcommand{\ee}{\end{equation}}
\newcommand{\ba}{\begin{array}}
\newcommand{\ea}{\end{array}}
\newcommand{\bea}{\begin{eqnarray}}
\newcommand{\eea}{\end{eqnarray}}
\def\vphi{\varphi}
\def\vrho{\varrho}
\def\vtheta{\vartheta}
\begin{center}
{\Large{\bf Two exactly-solvable problems in one-dimensional
\\[3mm]
quantum mechanics on circle}}

\vspace{0.6cm}
{\bf L.G.Mardoyan$^{1}$, G.S.Pogosyan$^{1,2,3}$ and A.N.Sissakian$^{2}$}

\vspace{0.3cm}
$^{1}${\it International Center for Advanced Studies,
Yerevan State University, Yerevan, Armenia}

\vspace{0.3cm}
$^{2}${\it Laboratory of Theoretical Physics, Joint Institute for Nuclear
Research, \\
Dubna, 141980, Moscow Region, Russia}

\vspace{0.3cm}
$^{3}${\it Centro de Ciencias F\'{\i}sicas, Universidad Nacional Aut\'onoma
de M\'exico \\
Apartado Postal 48--3,  Cuernavaca, Morelos 62251, M\'exico}
\end{center}

\vspace{0.4cm}

\begin{center}
{\bf Abstract}
\end{center}

\vspace{0.3cm}
\noindent

In this note we establish a relation between two exactly-solvable
problems on circle, namely singular Coulomb and singular oscillator
systems.

\vspace{0.7cm}
{\bf 1}. In a recent paper \cite{KMP2} have constructed a series of
complex mappings $S_{2C}\rightarrow S_2$, $S_{4C}\rightarrow S_3$
and $S_{8C}\rightarrow S_5$, which extend to spherical geometry the
Levi-Civita, Kustaanheimo-Stiefel and Hurwitz transformations, well
known for Euclidean space. We have shown that these transformations
establish a correspondence between Coulomb and oscillator problems in
classical and quantum mechanics for dimensions (2,2), (3,4) and (5,8) on
the spheres. A detailed analysis of the real mapping on the curved space
has been done in \cite{NERPOG}. It was remarked that in the stereographic
projection the relation between Coulomb and oscillator problems functionally
coincide with the flat space Levi-Civita and Kustaanheimo-Stiefel
transformations. The relation between the quasiradial Schr\"odinger equations
for Coulomb and oscillator problems on the $n$-dimensional spheres and
one- and two-sheeted hyperboloids for $n\geq 2$ was found in the article
\cite{KMJP}.

The present note are devoted to two one-dimensional exactly-solvable
potentials on circle $S_1$: $s_0^2+s_1^2=R^2$
%===============================================================
\begin{eqnarray}
\label{POT-SOC1}
V^{so}({\vec s}) =  \frac{\omega^2 R^2}{2} \frac{s_1^2}{s_0^2} + \frac{1}{2}
\frac{k_1^2-\frac14}{s_1^2}, \qquad V^{c}({\vec s}) =  - \frac{\mu}{R}
\frac{s_0}{|s_1|} + \frac{1}{2}\frac{p^2-\frac14}{s_1^2},
\end{eqnarray}
%===============================================================
where $s_0, s_1$ are Cartesian coordinates in ambient Euclidean space $E_2$.

\vspace{0.5cm}

{\bf 2.} The Schr\"odinger equation describing the nonrelativistic quantum
motion on the circle $S_1$ in polar coordinate $\vphi \in [-\pi, \pi]$
%===============================================================
\begin{eqnarray*}
s_0 = R\cos\vphi, \qquad s_1 = R\sin\vphi
\end{eqnarray*}
%===============================================================
has the following form ($\hbar = m =1$)
%===============================================================
\begin{eqnarray}
\label{SCH1-0}
\frac{d^2 \Psi}{d\vphi^2} + 2R^2[E - V(\vphi)]\Psi = 0
\end{eqnarray}
%===============================================================
Substituting the singular oscillator potential $V^{so}(\vphi)$ in
eq.~(\ref{SCH1-0}), we obtain P\"oschl-Teller-type equation
%===============================================================
\begin{eqnarray}
\label{OSCIL1}
\frac{d^2 \Psi}{d\vphi^2} + \left[\epsilon -
\frac{k_0^2-\frac14}{\cos^2\vphi} -
\frac{k_1^2-\frac14}{\sin^2{\vphi}}\right]\, \Psi = 0
\end{eqnarray}
%===============================================================
where $\epsilon = {2R^2E} + \omega^2 R^4$ and $k_0^2 = \omega^
2 R^4 + \frac14$. The regular (at points $\vphi = 0$ and $\pi/2$) solution
of above equation maybe chosen in following form \cite{FLUG}
%===============================================================
\begin{eqnarray}
\label{OSCFUN1}
\Psi_{n}(\vphi) = \sqrt{\frac{2(2n+k_0\pm k_1+1)\Gamma(n+k_0+k_1+1)
\Gamma(n \pm k_1+1)}{R\, [\Gamma(1 \pm k_1)]^2 \Gamma(n+k_0+1)(n)!}}
\nonumber\\[3mm]
(\sin\vphi)^{\frac12 \pm k_1} \, (\cos\vphi)^{\frac12 + k_0} \,
{_2F_1}(-n, n+k_0 \pm k_1+1; \, 1 \pm k_1; \, \sin^2\vphi),
\end{eqnarray}
%===============================================================
with
%===============================================================
\begin{eqnarray}
\label{ENER1}
\epsilon = (2n \pm k_1+k_0+1)^2, \qquad n=0,1,2.....
\end{eqnarray}
%===============================================================
Then the energy spectrum of the one-dimensional singular oscillator
is given by
%===============================================================
\begin{eqnarray}
\label{ENER11}
E_{n}(R) = \frac{1}{2R^2}\left[(2n \pm k_1+\frac12)^2 +
(2k_0+1)(2n \pm k_1+1)\right].
\end{eqnarray}
%===============================================================
Let us remark that the wave-functions have been normalized in the
domain $[0, \pi/2]$. The positive sign at the $k_1$ has to taken
whenever $k_1>\frac12$, i.e. the additional term to oscillator potential
is repulsive at the origin and the motion take place only in domain
$\vphi \in [0, \pi/2]$. If $0< |k_1| \leq \frac12$, i.e. the additional
term is attractive at the origin, both the positive and negative sign must
be taken into account in the solution. The motion in this case take place
in $\vphi \in [-\pi/2, \pi/2]$. This is indicated by the notion $\pm k_1$,
in the formulas.

\vspace{0.3cm}
\noindent
{\bf 3.} Let us to write the Schr\"odinger equation (\ref{SCH1-0}) for
singular Coulomb potential $V^{c}({\vec s})$
%===============================================================
\begin{eqnarray}
\label{KC1}
\frac{d^2 \Psi}{d\vphi^2} + \left(2R^2E + 2\mu R \cot|\vphi|+
\frac{p^2-\frac14}{\sin^2\vphi}\right) \, \Psi = 0.
\end{eqnarray}
%===============================================================
First we will consider the region $\vphi\in [0, \pi]$. We make now a
transformation to the new variable $\theta \in [0, \frac{\pi}{2}]$
%===============================================================
\begin{eqnarray}
\label{TRANS1}
e^{i\vphi} = \cos\theta,
\end{eqnarray}
%===============================================================
which is possible if we continue the variable $\vphi$ in the complex
domain: Re\ $ \vphi = 0$, $0\leq$\ Im\ $\vphi < \infty$ (see Fig. 1).
We complexify also the coupling constant $\mu$ by putting $k=i\mu$ such
that
%===============================================================
\begin{eqnarray}
\label{TRANS2}
\mu \cot\vphi =  k (1- 2 \sin^{-2}\theta).
\end{eqnarray}
%===============================================================
As result we obtain the equation
%===============================================================
\begin{eqnarray}
\label{COOS1}
\frac{d^2 W}{d\theta^2} + \left[\epsilon - \frac{k_0^2 - \frac{1}{4}}
{\cos^2\theta} - \frac{k_1^2 - \frac{1}{4}}{\sin^2\theta} \right]\, W = 0,
\end{eqnarray}
%===============================================================
where $W(\theta) = ({\cot\theta})^{\frac12} Z(\theta)$ and
%===============================================================
\begin{eqnarray}
\label{ENER2}
\epsilon = 2R^2 E + 2kR, \quad k_0^2 = 2R^2 E - 2kR,
\quad k_1^2 = 2 - 4p^2.
\end{eqnarray}
%===============================================================
From the above equation we see that, up to the substitution (\ref{ENER2})
the equation (\ref{COOS1}) for one-dimensional singular Coulomb problem
(with restriction $0 < p^2 \leq \frac12$) coincides to the one-dimensional
singular oscillator equation (\ref{OSCIL1}).

\vspace{7mm}
%=================================================================
\unitlength=1.00mm
\special{em:linewidth 0.4pt}
\linethickness{0.4pt}
\begin{picture}(102.00,126.60)
\put(14.00,44.00){\makebox(0,0)[cc]{0}}
\put(12.00,120.00){\makebox(0,0)[cc]{Im $\vphi$}}
\put(102.00,42.00){\makebox(0,0)[cc]{Re $\vphi$}}
\put(98.00,50.00){\line(-1,0){91.00}}
\put(65.00,50.00){\line(0,1){67.00}}
\put(65.00,43.00){\makebox(0,0)[cc]{$\pi$}}
\put(65.00,119.00){\line(-1,0){4.00}}
\put(59.00,119.00){\line(-1,0){4.00}}
\put(52.00,119.00){\line(-1,0){5.00}}
\put(44.00,119.00){\line(-1,0){5.00}}
\put(36.00,119.00){\line(-1,0){5.00}}
\put(28.00,119.00){\line(-1,0){5.00}}
\put(41.00,86.00){\makebox(0,0)[cc]{{\bf G}}}
\put(74.00,86.00){\makebox(0,0)[cc]{G}}
\put(96.00,50.00){\vector(1,0){2.00}}
\put(39.00,50.00){\vector(1,0){3.00}}
\put(65.00,84.00){\vector(0,1){1.00}}
\put(44.00,119.00){\vector(-1,0){2.00}}
\put(20.00,86.00){\vector(0,-1){2.00}}
\put(-5.00,22.00){\makebox(0,0)[lc]{{\bf Figure 1:}\ \
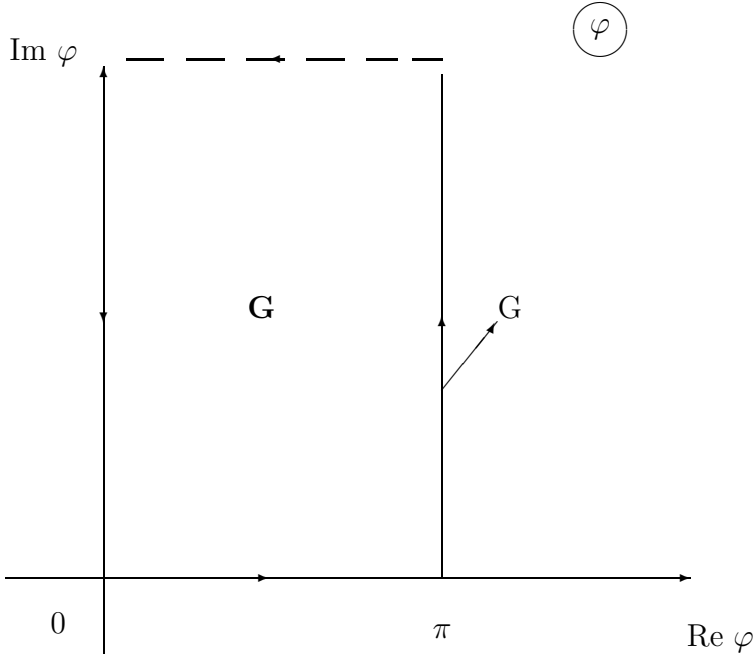
Domain {\bf G} =\{ $0\leq$ Re $\vphi\leq \pi$; $0\leq$ Im $\vphi< \infty$\}
on the complex plane of $\vphi$.}}
\put(20.00,40.00){\line(0,1){78.00}}
\put(86.00,123.00){\circle{7.21}}
\put(86.00,123.00){\makebox(0,0)[cc]{$\vphi$}}
\put(65.00,75.00){\line(4,5){7.33}}
\put(71.00,82.00){\vector(1,2){1.00}}
\put(20.00,118.00){\vector(0,1){0.00}}
\end{picture}
%=================================================================
\vspace{-9mm}

\noindent
The regular, for $\theta\in [0,\pi/2]$ and $k_0 \geq 1/4$, solution of
this equation according to (\ref{OSCFUN1}) is
%===============================================================
\begin{eqnarray}
\label{OSCFUN2}
\Psi(\theta) = \frac{W(\theta)}{\sqrt{\cot\theta}}
=C_{n} \, (\sin\theta)^{1\pm k_1} \, (\cos\theta)^{k_0}
\, {_2F_1}(-n, n + k_0 \pm k_1 + 1; \, 1 \pm k_1; \, \sin^2\theta)
\end{eqnarray}
%===============================================================
where $C_{n}$ is the normalization constant. To compute the constant
$C_{n}$ we require that the wave function (\ref{OSCFUN2}) satisfy the
condition
%===============================================================
\begin{eqnarray}
R\,\int_{0}^{\pi}\Psi_{n} \, \Psi_{n}^{\diamond} \, d\vphi = \frac12,
\end{eqnarray}
%===============================================================
where the symbol ``${\diamond}$'' means the complex conjugate together
with the inversion $\vphi\rightarrow -\vphi$, i.e.
$\Psi^{\diamond}(\vphi) = \Psi^{*}(-\vphi)$. [We choose the scalar product
as $\Psi^{\diamond}$ because for $\vphi\in${\bf G} and real $\mu$, and
$\epsilon$ the function $\Psi^{\diamond}(\vphi)$ also belongs to the solution
space of (\ref{KC1})]. By analogy to the work \cite{KMP2,KMJP}, we consider
the integral over contour $G$ in the complex plane of variable $\vphi$
(see Fig.1)
%================================================================
\bea
\label{NOR2}
\oint \Psi_{n}(\vphi) \, \Psi_{n}^{\diamond}(\vphi)
d\vphi &=& \int_{0}^{\pi} \Psi_{n}(\vphi) \, \Psi_{n}^{\diamond}(\vphi)
d\vphi + \int_{\pi}^{\pi+i\infty} \Psi_{n}(\vphi) \, \Psi_{n}^{\diamond}
(\vphi) d\vphi \nonumber\\[2mm]
&+& \int_{\pi+i\infty}^{i\infty} \Psi_{n}(\vphi) \, \Psi_{n}^{\diamond}
(\vphi) d\vphi + \int_{i\infty}^0 \Psi_{n}(\vphi) \, \Psi_{n}^{\diamond}
(\vphi) d\vphi.
\eea
%================================================================
Using the facts that the integrand vanishes as $e^{2i k_0\vphi}$ and that
$\Psi_{n}(\vphi)$ is regular in the domain {\bf G} (see Fig.1), then
according to the Cauchy theorem we have
%================================================================
\bea
\label{NOR3}
\int_{0}^{\pi} \Psi_{n}(\vphi) \, \Psi_{n}^{\diamond}(\vphi) d\vphi
= \left(1-e^{2i\pi k_0}\right)\, \int_0^{i\infty} \Psi_{n}(\vphi) \,
\Psi_{n}^{\diamond}(\vphi) d\vphi.
\eea
%================================================================
Making the substitution (\ref{TRANS1}) in the right integral of
eq. (\ref{NOR3}), we find
%================================================================
\bea
\int_{0}^{\pi} \Psi_{n}(\vphi) \, \Psi_{n}^{\diamond}(\vphi) d\vphi
= i\left(1-e^{2i\pi k_0}\right)\, \int_{0}^{\frac{\pi}{2}}
\left[\Psi_{n}\right] \tan\theta \, d\theta.
\eea
%================================================================
and after integration over the angle $\theta$ we finally get \cite{BER}
%===============================================================
\bea
\label{SSS1}
C_{n} = \sqrt{\frac{(-ik_0)(2n+k_0 \pm k_1+1)\Gamma(n+1 \pm k_1)
\Gamma(n+k_0 \pm k_1+1)} {R [1-e^{2i\pi k_0}](2n\pm k_1+1) n!
[\Gamma(1 \pm k_1)]^2 \Gamma(n+k_0+1)}}.
\eea
%================================================================

Let us now consider two most interesting cases.

\noindent
{\bf (i)}. The case when $p^2=\frac14$. Then the duality transformation
(\ref{TRANS1}) establish the connection between pure Coulomb problem and
singular oscillator with $k_1^2=1$. Comparing the eqs.~(\ref{ENER1}) with
(\ref{ENER2}) and putting $k=i\mu$, we get
%===============================================================
\begin{eqnarray}
\label{OMEGA1}
k_0 = - (n + 1) + i\sigma, \qquad \sigma = \frac{\mu R}{n + 1}
\end{eqnarray}
%===============================================================
and for energy spectrum
%===============================================================
\begin{eqnarray}
\label{ENERGY1}
E_n (R) = \frac{(n + 1)^2}{2R^2} - \frac{\mu^2}{2(n+1)^2},
\qquad n= 0,1,2,....
\end{eqnarray}
%===============================================================
Returning to the variable $\vphi$, we obtain that wave function at
$0\leq\phi\leq \pi$ has the form
%=================================================================
\bea
\label{SS22}
\Psi_{n\sigma}(\varphi) = C_{n}(\sigma) \,\, \sin\varphi \,\,
e^{-i\varphi(n-i\sigma)} \,\, {_2F_1}(-n, \, 1+i\sigma; \, 2; \,
1-e^{2i\varphi})\, ,
\eea
%=================================================================
where the normalization constant $C_{n}(\sigma)$ is
%=================================================================
\bea
\label{NSS22}
C_{n}(\sigma) = e^{\frac{\sigma\pi}{2}}\, |\Gamma(1+i\sigma)| \,
\sqrt{\frac{(n+1)^2+\sigma^2}{\pi R}}.
\eea
%=================================================================
The wave function in the region $-\pi \leq \varphi < 0$ ($s_1 < 0$)
maybe determined from the eq.~(\ref{SS22}) by the reflection
$\varphi \to -\varphi$. Therefore the general solution of the Schr\"odinger
equation for $\varphi \in [-\pi, \pi]$ can be presented in form of even
and odd function
%=========================(16)===============================
\bea
\label{S-S1}
\Psi_{n\sigma}^{(+)}(\varphi) &=& C_{n}(\sigma) \,\, \sin |\varphi|
e^{-i(n-i\sigma)|\varphi|} F\left(-n, 1+i\sigma; 2; 1-e^{2i|\varphi|}\right),
\nonumber \\[2mm]
\Psi_{n\sigma}^{(-)}(\varphi) &=& C_{n}(\sigma) \,\, \sin \varphi \,\,
e^{-i(n-i\sigma)|\varphi|}
F\left(-n, 1+i\sigma; 2; 1-e^{2i|\varphi|}\right). \nonumber
\eea
%============================================================
Thus by using the relation between Coulomb and singular oscillator
systems we have constructed the wave functions and energy spectrum
for a Coulomb system on the one-dimen\-si\-onal sphere.

\noindent
{\bf (ii)}. Choosing now $k_1^2=\frac14$ or the same $p^2 - \frac14 =
\frac{3}{16}$. In this case the centrifugal potential term is repulsive
at the origin for singular Coulomb systems and the motion take place only
in one of the domains $\vphi > 0$ or $\vphi < 0$. For oscillator system
this term equal zero and therefore the duality transformation (\ref{TRANS1})
connect singular Coulomb and pure oscillator systems.

Let us introduce the quantity $\nu$ which takes two values
$\nu = \frac14$ and $\nu=\frac34$. Making all calculations by analogy
to previous case, it is easy to obtain the energy spectrum
%===============================================================
\begin{eqnarray}
\label{ENERGY1-0}
E_n^{\nu}(R) = \frac{(n + \nu)^2}{2R^2} - \frac{\mu^2}{2(n+\nu)^2},
\qquad n= 0,1,2,....
\end{eqnarray}
%===============================================================
and wave functions
%=================================================================
\bea
\label{SS22-0}
\Psi_{n\sigma}^{\nu}(\varphi) &=& e^{\frac{\sigma\pi}{2}}\, 2^\nu \,
\frac{|\Gamma(\nu+i\sigma)|}{\Gamma(2\nu)} \,
\sqrt{\frac{[(n+\nu)^2+\sigma^2]\Gamma(n+2\nu)}{4\pi R (n+\nu)n!}}.
\nonumber\\[2mm]
&\times&(\sin\varphi)^{\nu} \,\, e^{-i\varphi(n-i\sigma)}
\,\, {_2F_1}(-n, \, \nu+i\sigma; \, 2\nu; \, 1-e^{2i\varphi})\, ,\nonumber
\eea
%=================================================================
where $\sigma= \mu R/(n+\nu)$.

In the contraction limit $R\to \infty$, $\vphi \to 0$ and
$R \vphi \sim x$ - fixed, we see that
%=================================================================
\bea
\label{SS22-11}
\epsilon_n = \lim_{R\rightarrow\infty} E_n^{\nu}(R) = -
\frac{\mu^2}{2(n+\nu)^2}, \qquad n=0,1,...
\eea
%=================================================================
and
%=================================================================
\bea
\label{SS22-12}
\Phi_n^{\nu}(y)= \lim_{\scriptstyle R\rightarrow\infty\atop\scriptstyle
\varphi\rightarrow 0} \Psi_{n\sigma}^{\nu}(\vphi) =
\frac{\sqrt{\mu}}{\Gamma(2\nu)} \frac{1}{(n+\nu)} \,
\sqrt{\frac{\Gamma(n+2\nu)}{2n!}}\, y^{\nu}\, e^{- \frac{|y|}{2}}\,
{_1F_1}(-n; \, 2\nu; \, y),
\eea
%=================================================================
where $y = {2\mu x}/{(n+\nu)}$. Formulas (\ref{SS22-11}) and
(\ref{SS22-12}) coincides with the formulas for energy levels $\epsilon_n$
and up to the factor $\sqrt{2}$ for wave functions $\Phi_n^{\nu}(y)$ for
two type one-dimensional Coulomb anyons with $\nu=\frac14$ and $\nu=\frac34$
respectively \cite{TER-ANT}.

\vspace{0.5cm}

\noindent
{\bf Acknowledgments}

The research of L.G.M. and G.S.P. was partially supported by the NATO 
Collaborative Linkage Grant No. 978431 and ANSEF Grant No. PS81.

\end{document}